\documentclass[runningheads]{llncs}
\usepackage{graphicx}
\usepackage{version}

\usepackage{listings}

\usepackage{listings, xcolor}

\definecolor{verylightgray}{rgb}{.97,.97,.97}

\lstdefinelanguage{Solidity}{
	keywords=[1]{anonymous, assembly, assert, balance, break, call, callcode, case, catch, class, constant, continue, contract, debugger, default, delegatecall, delete, do, else, event, export, external, false, finally, for, function, gas, if, implements, import, in, indexed, instanceof, interface, internal, is, length, library, log0, log1, log2, log3, log4, memory, modifier, new, payable, pragma, private, protected, public, pure, push, require, return, returns, revert, selfdestruct, send, storage, struct, suicide, super, switch, then, this, throw, transfer, true, try, typeof, using, value, view, while, with, addmod, ecrecover, keccak256, mulmod, ripemd160, sha256, sha3}, 
	keywordstyle=[1]\color{blue}\bfseries,
	keywords=[2]{address, bool, byte, bytes, bytes1, bytes2, bytes3, bytes4, bytes5, bytes6, bytes7, bytes8, bytes9, bytes10, bytes11, bytes12, bytes13, bytes14, bytes15, bytes16, bytes17, bytes18, bytes19, bytes20, bytes21, bytes22, bytes23, bytes24, bytes25, bytes26, bytes27, bytes28, bytes29, bytes30, bytes31, bytes32, enum, int, int8, int16, int24, int32, int40, int48, int56, int64, int72, int80, int88, int96, int104, int112, int120, int128, int136, int144, int152, int160, int168, int176, int184, int192, int200, int208, int216, int224, int232, int240, int248, int256, mapping, string, uint, uint8, uint16, uint24, uint32, uint40, uint48, uint56, uint64, uint72, uint80, uint88, uint96, uint104, uint112, uint120, uint128, uint136, uint144, uint152, uint160, uint168, uint176, uint184, uint192, uint200, uint208, uint216, uint224, uint232, uint240, uint248, uint256, var, void, ether, finney, szabo, wei, days, hours, minutes, seconds, weeks, years},	
	keywordstyle=[2]\color{teal}\bfseries,
	keywords=[3]{block, blockhash, coinbase, difficulty, gaslimit, number, timestamp, msg, data, gas, sender, sig, value, now, tx, gasprice, origin},	
	keywordstyle=[3]\color{violet}\bfseries,
	identifierstyle=\color{black},
	sensitive=false,
	comment=[l]{//},
	morecomment=[s]{/*}{*/},
	commentstyle=\color{gray}\ttfamily,
	stringstyle=\color{red}\ttfamily,
	morestring=[b]',
	morestring=[b]"
}

\begin{document}
\title{Decentralized Autonomous Organizations for Tax Credit's Tracking}
\titlerunning{DAOs for Tax Credit's Tracking}
%
\author {Giovanni De Gasperis\inst{1}\orcidID{0000-0001-9521-4711} 
\and Sante Dino Facchini\inst{1}\orcidID{0000-0002-2009-5209} 
\and Alessio Susco\inst{1}}
\authorrunning{G. De Gasperis et al.}
%
\institute{DISIM, Università degli Studi dell’Aquila, Via Vetoio, L’Aquila IT67100, Italy }
%
\maketitle              
\begin{abstract}
Tax credit stimulus and fiscal bonuses had a very important impact on Italian economy in the last decade. Along with a huge expansion in constructions a relevant increase in scams and frauds has come too. The aim of this article is to design a possible system to track and control the whole tax credit process from its generation to its redeem through a Decentralized Autonomous Organization architecture enriched with a Multi Agent Systems to implement controllers.


\end{abstract}

\section{Introduction: Tax credit stimulus in Italian law}
\textbf{A brief history of tax credit measures in Italy}. Italian energy policies and laws have a deep root into European Union guidelines, these mainly aim to improve the performance and to reduce the wastefulness of households energy production systems. During the last years the debate around energy efficiency rose in importance and is now central to determine the Next Generation EU funding policies,
As a EU Member, Italy has taken steps to encourage energy efficiency increase of residential buildings. A program of tax deductions and fiscal bonuses, originally introduced in 1997 (art.1 Law 449/1997) \footnote{https://www.gazzettaufficiale.it/eli/id/1998/01/28/098A0239/sg} were relaunched since 2011 (art.4 DL 201/2011)\footnote{https://www.gazzettaufficiale.it/eli/gu/2011/12/06/284/so/251/sg/pdf} and renewed over the years till 2019. Finally during 2020 the actual system of 110 percent Superbonus tax credit has been introduced (Art. 119 DL Rilancio 34/2020)\footnote{https://www.gazzettaufficiale.it/eli/id/2020/05/19/20G00052/sg}. For the first time the lawmaker set up a system where energy efficiency and structural works are at no cost for residential properties. 

\textbf{The actual legislation and the invoice discount}. Superbonus 110, unlike other tax allowances for energy efficiency measures and reduction of seismic risks of buildings, provides for a higher-than-investment rate of deduction as well as a different way of allowances claim.  
In fact for each 100 euros spent on renovation works, the buyer will mature a 110 euros tax deduction to be used over the next 5 years and divided into 5 equal annual instalments.
Since the annual deduction is discounted from the taxpayer's gross tax (IRPEF), it is therefore recoverable within the limit of such amount and  cannot be carried forward or claimed back. Any deduction in excess of the gross tax would be lost.
The legislator has thus provided two alternative solutions to the direct deduction: a discount on the invoice received for the works or assignment/selling of the matured credit to a third party.
The option considered and modeled in this paper is the Invoice Discount as it is the most diffused solution and, differently from direct selling or direct deduction, requires credits to be tracked and managed as the benefits may be transferred through many subjects during the generation and redeem process. 
It consists of a contribution in the form of a discount on the invoice, up to an amount equal to the total including VAT, to be requested to the construction companies. It in turn may monetize it by either using it in compensation for future taxes to pay or by selling such right to third parties (in most cases Financial Institutions). The critical features of invoice discount mechanism are the following: Customers have no disbursement for the restoration works; fiscal and technical visas are required to certify credits matured from the discounted invoices; Suppliers of goods and services are financially exposed, as do not receive cash payment for their invoices but accrues a tax credit equal to the deduction matured by the Costumer/Taxpayer and Suppliers may in turn assign/sell the credit to other parties.
Another assumption we make in this paper is that the Costumers are using a General Contractor to manage the whole process of renovation on the construction yard and connected paperworks. This is both in order to simplify the level of the model (which in turn can be easily extended to a multi supplier situation) and to describe the most diffused scenario.

\begin{figure}
\centering
\includegraphics[width=\textwidth]{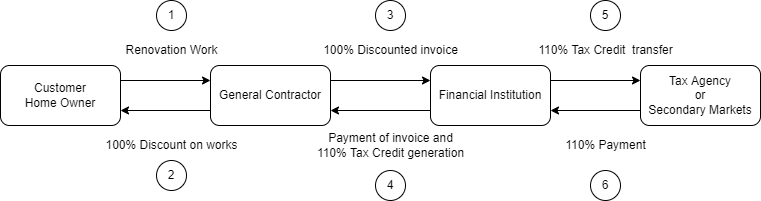}
\caption{Invoice Discount Tax Credit generation.} \label{fig1}
\end{figure}
The General Contractor will be the only subject dealing with the Costumer, he will take care of coordinating and paying all the sub-contractors involved in the renovation works as well as all professionals and technicians required both for technical and financial projects and certifications. As a result the discounted invoices will be raised only by the General Contractor which in turn will pay cash all the Suppliers and Professionals aforesaid, thus will be the only subject entitled to get the tax credit matured by the Costumer.

\textbf{Tax fraud mechanism and credit tracking}. In such an environment is of paramount importance being able to track the history of credits as well as the identity of all players involved in the process. Identity scams and credit assignment to fraudster recipients are in fact the main risks involved in Superbonus 110 process. Another risk, not connected necessary to willing to fraud, is represented by incorrect credit assignment where for example a client asks for a bigger amount than owed or more properties than allowed. In all of these situation back tracking of each euro of tax credit assigned is of key importance.

\textbf{A brief introduction to Distributed Autonomous Organizations}. The idea of Distributed Autonomous Organizations (DAOs) has been outlined for the first time during the late 90s and was connected to the application of Multi-agent Systems to intelligent home sensors \cite{Dilger}. With the introduction of the Digital Autonomous Corporations concept (DACs) after 2008, a first transposition of a real company on the blockchain was defined; taking advantage from the introduction of Ethereum instruments like tokenization of shares, fully automated and incorruptible smart contracts along with transparent transactions register were finally applicable to the real governance process of a company \cite{Buterin}.
The actual DAOs paradigm is an evolution of the DACs one, on the technical perspective it refers to a system able to model an organization through the deployment of a set of interacting smart contracts upon a blockchain network. It also inherit the properties of the underlying layers such as lack of centralized control, security through cryptographic keys and self-execution of smart contracts. All these properties make DAOs a reliable paradigm to operate a virtual organization modeling all functionalities and procedures of a real one.

\section{Main purpose: DAOs for Tax Credit's tracking}
\textbf{Roles, Actors and entities}. The aim of this paper is to investigate the application of DAO paradigm to manage the procedures of Superbonus 110 evaluating a possible integration with a MAS based intelligent control system. The first step is to identify the environment and the entities involved in the process, we identified to start two main areas: the investors group which includes all actors getting financial burdens and benefits in the process, and the operators group that in turns aggregate all actors with on-the-job interests in the Superbonus. Some actors may take both groups properties as per Table \ref{tab:my-table}.

\begin{table}[]
\begin{tabular}{|p{0.23\textwidth}|p{0.24\textwidth}|p{0.53\textwidth}|}
\hline
\multicolumn{1}{|c|}{\textbf{Group}} & \multicolumn{1}{c|}{\textbf{Actors}} & \multicolumn{1}{c|}{\textbf{Role and Actions}} \\ \hline
Investors                            & Investor                               & Invest money to buy credits and benefits from its selling          \\ \hline
Investors-Operators              & Financial Institution                  & Banks that buy/sell credits and lend money to Operators                                               \\ \hline
Investors-Operators              & Customer                  & House owner Sells/Transfers tax creditsfrom works                                               \\ \hline
Operators                            & General Contractor                     & Manages works, gets credits from Customers, sells to Financial Institutions                                               \\ \hline
Operators                            & Sub-contractor                         & Hired and paid by General Contractor                                               \\ \hline
Operators                            & Supplier                               & Hired and paid by General Contractor                                               \\ \hline
Operators                            & Design Architect                       & Hired and paid by General Contractor                                               \\ \hline
Operators                            & Tax Auditor                            & Hired and paid by General Contractor                                               \\ \hline
\end{tabular}
\caption{Users and Roles}
\label{tab:my-table}
\end{table}
 \textbf{Double DAOs architecture}. The software and logic architecture proposed to implement the interaction between such actors is based on the separation of investors' world by the operators' one. This is due to consent Financial Institutions to raise fund from clients and investors and tokenize such assets, the process is made assuming a simplified model where banks receive transfers of fiat on their accounts and mint the correspondent value of tokens on the first DAO, the Investors one.
 Another key feature is to guarantee that money anticipation performed by financial institutions and the correspondent underlying credit generation is backed by a fiat asset and is thus repayable when the credits' chain is closed.  In brief the first DAO system, named Superbonus Investors, is meant to represent a simplified investing fund where participants get tokens proportionally to their investment, keep them blocked until the end of the fund duration and get a reward proportional to their initial quota at the end. Here we have Non Fungible Tokens (NFT) that are minted upon fiat deposit or credit selling by the Financial Institution and freezed or unfreezed when a guarantee from Operators DAO is requested or released. The architecture can be seen as a Distributed Trust and Reputation Management Systems (DTRMS) environment implemented through two DAOs.
 
\begin{figure}[!ht]
\centering
\includegraphics[width=\textwidth]{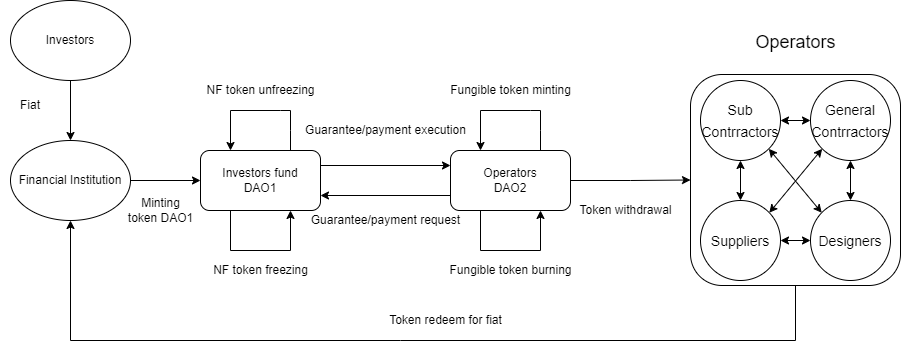}
\caption{DAOs Interaction Diagram.} \label{dao}
\end{figure}
 
 The second DAO is basically a marketplace where the Financial Institution distribute fungible tokens to General Contractors, these are minted and burned in order to satisfy credit request from Operators and spending for buying good and services. Each minting on Operators DAO correspond to a token freezing of the same amount on first DAO in order to guarantee the coverage. The tokens are redeemable by the  Financial Institution that is also warrantor of the whole system. Once second DAO tokens are redeemed, first DAO NFTs are unfreezed. Each fungible token minted here is coupled via a unique code to the tax credit generated by that spending of money.

\textbf{Aragon as developing platform}. The framework chosen to develop our demonstrator is Aragon \footnote{https://aragon.org/}. This is a second level platform based on the Ethereum network \footnote{https://ethereum.org/} that is natively designed to model government systems of public and private entities, it is easy to use and allows to deploy test nets for application at very cheap costs using Ethereum's test nets such as Rinkeby \footnote{https://www.rinkeby.io/}. Another feature of the Aragon framework is that completely manages the layers necessary for communication between the dApp and the Ethereum Virtual Machine (EVM); in this way it is possible to concentrate on the development of business logic and the interaction with external smart contracts via the Agent module of the framework.
The Aragon stack is mainly composed of: an Operative system environment where applications are abstracted from the underlying layer, a Packet Manager to distribute different versions of the software, APIs to manage requests on transactions, status and software without depending on a centralised service and a toolkit for the development of the dApps user interface.

\textbf{Internal DAOs Smart contracts}. The model described in the previous section is implemented through several smart contracts, deployed for this demonstration on Rinkeby test net and integrated with the Aragon management platform. Contracts runs on the Ethereum Virtual Machine and are written with Solidity \footnote{https://soliditylang.org/} an object-oriented language specifically meant for contract contracts implementation. The choice of such stiff language is because we want the execution to be strictly what is supposed to be without possibility of . A solidity example of Operators token minting function is illustrated below:

\begin{lstlisting}[language=Solidity]
function MintOperator(address _to, uint256 _amount) external auth(MINT_OPERATOR_ROLE) {
    //Unfreezed Investor plafond must have at least the same amount of Operator Tokens that we want to mint
    require(getUnfreezedInvestorBalance() >= _amount);

    //Burn Unfreezed Investor Token first
    BurnUnfreezedInvestorToken(_amount);

    //Mint same amount of Freezed Investor Token with the Operator Token's LinkID
    MintFreezedInvestorToken(_amount);

    //Then Mint real Fungible Operator Tokens and assign them to the target owner
    MintOperatorToken(_to, _amount);
}
\end{lstlisting}

\begin{figure}[!ht]
\centering
\includegraphics[width=\textwidth]{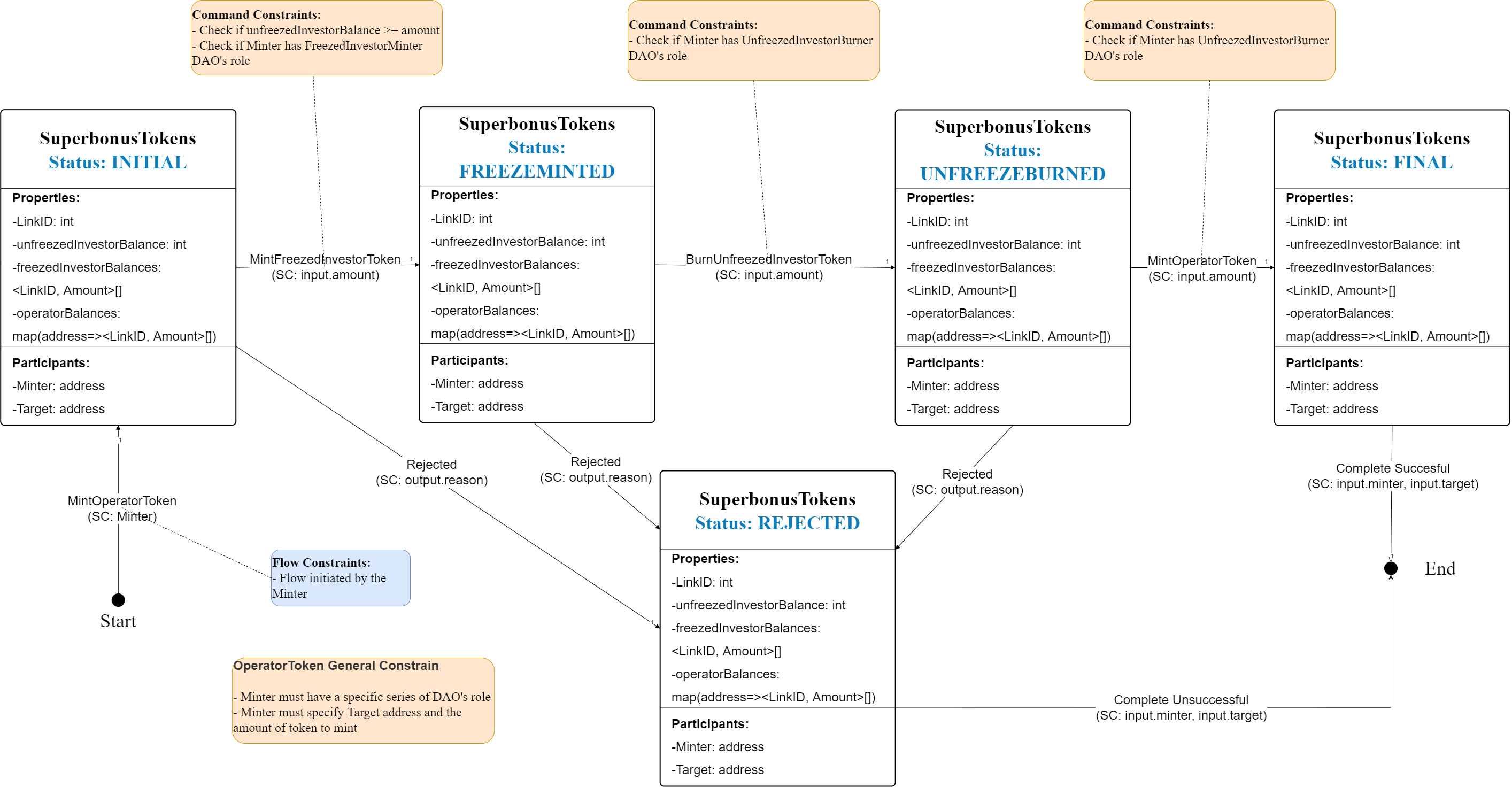}
\caption{Smart Contract View.} \label{smart_contract}
\end{figure}

A representation of Smart contract's states along with global and local constrain, data structures and actors involved is instead reported in Figure \ref{smart_contract}

\begin{figure}[!ht]
\centering
\includegraphics[width=\textwidth]{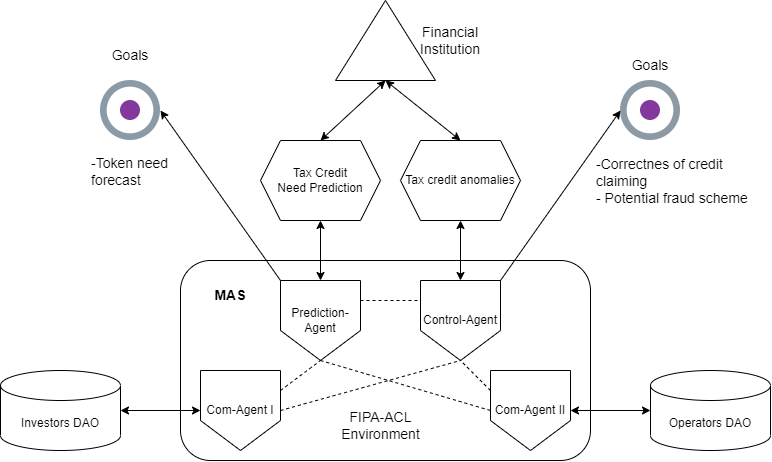}
\caption{MAS architecture.} \label{mas}
\end{figure}

\textbf{Integration of Multi Agent Systems}. We have integrated a simple module composed of 4 Agents in order to control main aspects of the Investors an Operators DAOs and suggest the future behaviour of the Financial Institutes in terms of token needs of the system . Com-agents manage the data exchange with the DAOs, the Prediction-agent forecasts token needs on Operators DAO while Control-agent checks the correct behaviour of operators in claiming credits and points put potential fraud or incorrect schemes.
We used the MESSAGE/UML paradigm to model the agent interaction and behaviour\cite{Ozakaya} in Figure \ref{mas}
\section{A sample application (or demonstrator)}
The demonstrator we have implemented is meant to illustrate the benefits of introducing the DAOs and MAS to Superbonus 110 environment. The actual procedure has two weak point: the first is the difficulty to track the credits and the second the disbursement of anticipated cash both from Financial Institutions and General Contractors. In this way we'll focus on two main features of the software in order to simulate the substitutions of cash with Operators tokens and the tracking of the matured credits through a spreading tree. It will be possible to track the tokens coupled to tax credit and represent their path as a tree where the root is the first minting action.
Furthermore the MAS modules will point out potential fraud situations and forecast the token request path in a certain time lapse. 
\begin{figure}[!ht]
\centering
\includegraphics[width=\textwidth]{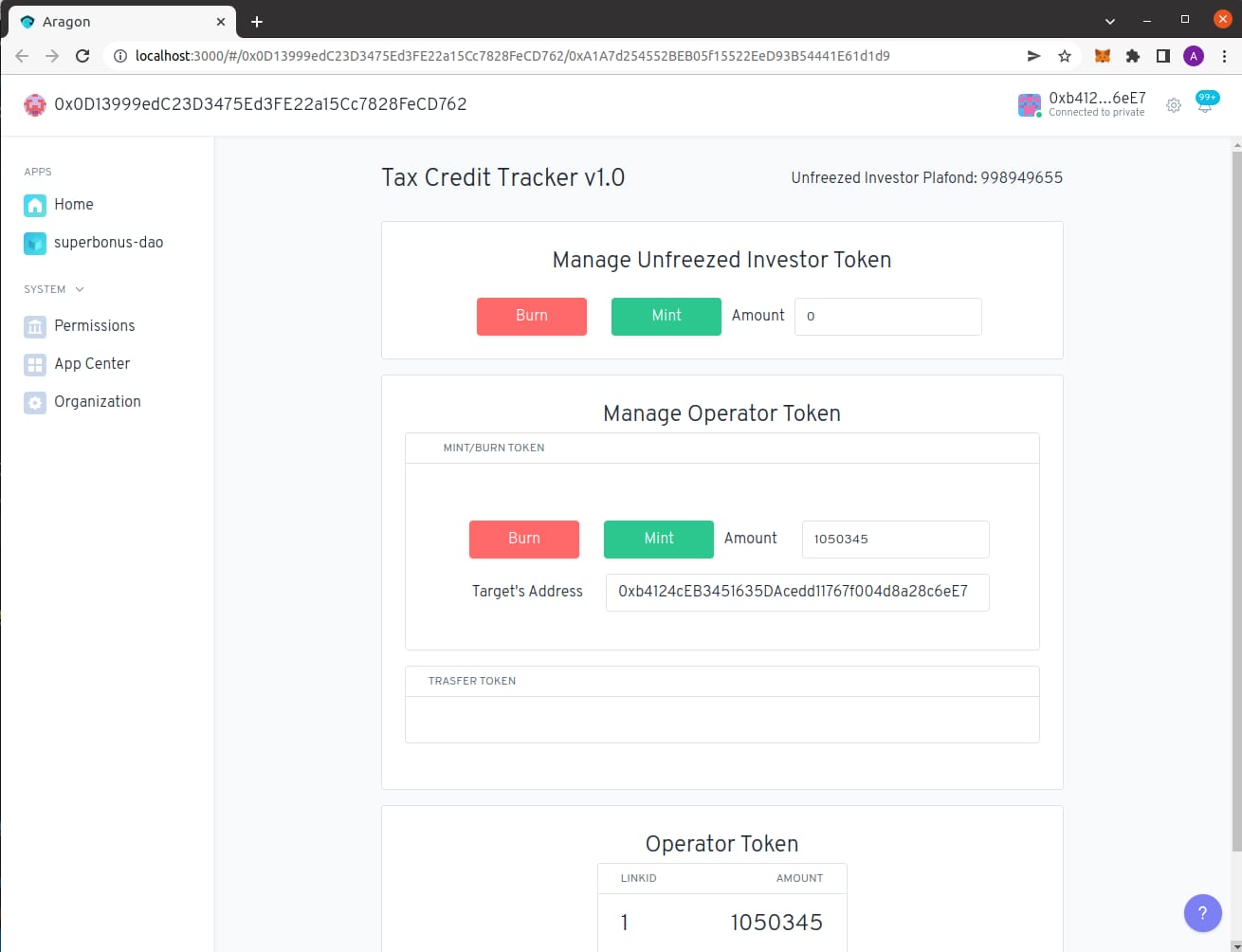}
\caption{Demonstrator User Interface.} \label{sw_ui}
\end{figure}
\section{Conclusions}
The goal of our demonstrator is to track the life cycle of tax credits generated by Superbonus 110 incentives scheme. The model used could be easily exported to other areas both in the institutional and private sectors. In particular public administrations may benefit of a reliable governance and voting system to provide services that requires tracking of items and documents (eg. Board resolutions, Certifications, etc)\cite{Chohan}. More in general every process where a tracking of some asset or document is required could benefit from the architecture of Blockchain in the environment of Distributed Trust and Reputation Management Systems (DTRMS)\cite{Bellini}. Another possible future development may consider the integration of smart contracts and agents with IoT solutions such as sensors to monitor physical quantities.


\begin{thebibliography}{8}
\bibitem{Dilger}
Dilger, Werner. Decentralized autonomous organization of the intelligent home according to the principle of the immune system. In: 1997 IEEE International Conference on Systems, Man, and Cybernetics. Computational Cybernetics and Simulation. IEEE, 1997. p. 351-356.

\bibitem{Buterin}
Buterin, Vitalik, et al. A next-generation smart contract and decentralized application platform. white paper, 2014, 3.37: 2.1.

\bibitem{Ozakaya}
Ozkaya, Mert. Analysing UML-based software modelling languages. Journal of Aeronautics and Space Technologies, 2018, 11.2: 119-134.

\bibitem{Chohan}
Chohan, Usman W. and Chohan, Usman W., The Decentralized Autonomous Organization and Governance Issues (December 4, 2017). Available at SSRN: https://ssrn.com/abstract=3082055 or http://dx.doi.org/10.2139/ssrn.3082055

\bibitem{Bellini}
E. Bellini, Y. Iraqi and E. Damiani, "Blockchain-Based Distributed Trust and Reputation Management Systems: A Survey," in IEEE Access, vol. 8, pp. 21127-21151, 2020, doi: 10.1109/ACCESS.2020.2969820.
\end{thebibliography}
\end{document}